\documentclass[runningheads]{llncs}
\usepackage[T1]{fontenc}
\usepackage{graphicx}
\usepackage{threeparttable}
\usepackage{multirow}
\usepackage{makecell}
\usepackage{bbding}
\usepackage{amsmath}
\usepackage{amsfonts}
\usepackage{booktabs}
\usepackage{svg}
\begin{document}
%
\title{DSNet: Disentangled Siamese Network with Neutral Calibration for Speech Emotion Recognition}
\titlerunning{DSNet}
\author{Chengxin Chen\inst{1,2}\orcidID{0000-0003-4510-3313} \and 
Pengyuan Zhang\inst{1,2}\orcidID{0000-0001-6838-5160}}
\authorrunning{C. Chen et al.}
%
\institute{Key Laboratory of Speech Acoustics and Content Understanding, Institute of Acoustics, Chinese Academy of Sciences, China \and
University of Chinese Academy of Sciences, China \\
\email{\{chenchengxin, zhangpengyuan\}@hccl.ioa.ac.cn}}
\maketitle              
\begin{abstract}
One persistent challenge in deep learning based speech emotion recognition (SER) is the unconscious encoding of emotion-irrelevant factors (\emph{e.g.}, speaker or phonetic variability), which limits the generalization of SER in practical use.
In this paper, we propose DSNet, a Disentangled Siamese Network with neutral calibration, to meet the demand for a more robust and explainable SER model.
Specifically, we introduce an orthogonal feature disentanglement module to explicitly project the high-level representation into two distinct subspaces.
Later, we propose a novel neutral calibration mechanism to encourage one subspace to capture sufficient emotion-irrelevant information.
In this way, the other one can better isolate and emphasize the emotion-relevant information within speech signals.
Experimental results on two popular benchmark datasets demonstrate the superiority of DSNet over various state-of-the-art methods for speaker-independent SER.

\end{abstract}

\section{Introduction}
Speech emotion recognition (SER), which aims to discern the emotional state from spoken language, has garnered significant interest in the field of affective computing~\cite{ser_review}. 
SER enables a more natural and harmonic human-machine interaction, fostering a variety of applications like user preference analysis, mental health care, and intelligent customer service. 
However, emotion is a relatively subtle variance hidden in speech and highly entangled with other acoustic factors, posing great challenges to the SER task.

Prior studies on SER were devoted to exploring the emotionally discriminative acoustic features in a hand-crafted way~\cite{handcraft1,handcraft2}.
In recent years, deep learning algorithms~\cite{sota1,sota2,sota3,sota4} have become the mainstream in SER, owning to the capacity of automatic feature abstraction from raw speech spectrogram.
Despite the advancements, SER models still suffer from severe performance degradation when tested on unseen data.
Due to the limited scale of available SER datasets, there often exist spurious correlations between emotion labels and (easy to learn) latent emotion-irrelevant attributes.
Consequently, models trained on such datasets are prone to learning the ``shortcuts'' and performing poorly when the correlations do not hold for the testing data.

One paradigm for handling this issue is to align such correlations by learning a shared feature space between training and testing data, also known as domain adaptation~\cite{dat1,dat2,dat3}.
For example, 
Luo \emph{et al.} introduced a non-negative matrix factorization based transfer subspace learning method~\cite{dat2} to minimize the discrepancy between the two distributions and exclude their individual components.
While appealing, these approaches typically require the unlabeled or partially labeled testing data to participate in training, and the models need to be retrained when adapted to new testing data.
Another research line aims to directly eliminate the influence of emotion-irrelevant biases on the training data by means of feature normalization~\cite{normalize1,normalize2,normalize3} or adversarial learning~\cite{debias-speaker,debias-speaker2,debias-speaker3}.
Unfortunately, most of these methods are task-oriented and require manual definition and labeling for a particular bias, making it difficult to look through all undesirable biases and remove them simultaneously.
More recently, researchers have explored several instance-oriented methods to reduce the distributional mismatch across different speakers~\cite{dsc,isnet}. 
The basic idea of these works is to model the relative variance between emotional and neutral speech of the same speaker.
It is still unclear, though, how to make the most of the internal correlation that exists between emotional and neutral speech.


Motivated by the above observations, we propose DSNet, a Disentangled Siamese Network with neutral calibration.
The core innovation of DSNet is to explicitly disentangle the emotion-relevant and -irrelevant components of given emotional speech under the supervision of homologous neutral speech.
Specifically, we assume that the deep representation extracted from neutral speech spectrogram contains all the latent biases of the same speaker, such as identity, gender, age, or accent, which can be regarded as a ``golden standard'' to guide the learning of the emotion-irrelevant representation.
Besides, we incorporate a combination of orthogonality and reconstruction constraints to ensure the complementarity of feature disentanglement.
Finally, the emotion-relevant representation is expected to facilitate the downstream classifier with better reliability and interpretability.
The main contributions of this paper can be summarized as follows:
\begin{itemize}
 \item We present DSNet, an instance-oriented approach that performs explicit feature disentanglement with a novel neutral calibration mechanism. To our best knowledge, this is the first study that integrates the ideas behind disentangled representation learning and Siamese neural network for SER.
 \item DSNet is a plug-and-play technique that can be easily integrated into different backbone architectures with a reasonable extra computational cost.
 \item Experiments on two benchmark datasets demonstrate that our proposed DSNet outperforms currently advanced speaker-independent SER algorithms in both within-corpus and cross-corpus settings.
\end{itemize}

The remainder of this paper is organized as follows: 
Section~\ref{Related Works} reviews the related works. 
Section~\ref{Methodology} introduces the overall workflow and detailed modules of our proposed method. 
Section~\ref{Experimental Datasets and Setup} describes the experimental datasets and setup.
Section~\ref{Experimental Results and Analysis} presents the experimental results and discussions.
Finally, Section~\ref{Conclusions} concludes this work and outlines future research directions.

\section{Related Works} \label{Related Works}
Disentangled Representation Learning (DRL) is an important paradigm for explainable and controllable machine learning by identifying and separating the hidden factors of the learned representation~\cite{drl}.
Early theoretical explorations of DRL showed significant promise in the field of computer vision, and they usually relied on certain model architectures, such as variational auto-encoder~\cite{vae} or generative adversarial network~\cite{gan}.
Recently, DRL has boosted various speech processing tasks, including speech synthesis~\cite{drl-tts}, voice conversion~\cite{drl-vc}, and speech enhancement~\cite{drl-se}.
These empirical methods mainly used task-specific pre-trained encoders to extract certain factors of speech and designed appropriate loss functions to ensure the disentanglement.
By contrast, this work proposes an instance-oriented DRL method to enhance the interpretability and reliability of SER.

Siamese Neural Network (SNN) is a type of models made up of two or more identical sub-networks that have the same architectures and weights assigned to the parameters~\cite{snn}.
Unlike typical neural networks, which learn to predict the correct labels by optimizing cross-entropy loss, SNN often measures the semantic similarity of inputs via contrastive loss~\cite{cl1,cl2}.
Specifically, contrastive loss is intended to promote discriminative yet semantically rich representations by drawing dissimilar samples apart and similar samples closer together.
Prior studies have explored the viability of using SNN for the speech classification tasks, such as emotion recognition~\cite{snn-er} and speaker recognition~\cite{snn-sr}.
In this work, we introduce SNN to calibrate the disentangled emotion-irrelevant components under the supervision of Siamese neutral representation.

\section{Methodology} \label{Methodology}
\begin{figure*}[htb]
  \centering
  \scalebox{0.9}
  {\includegraphics[width=\linewidth]{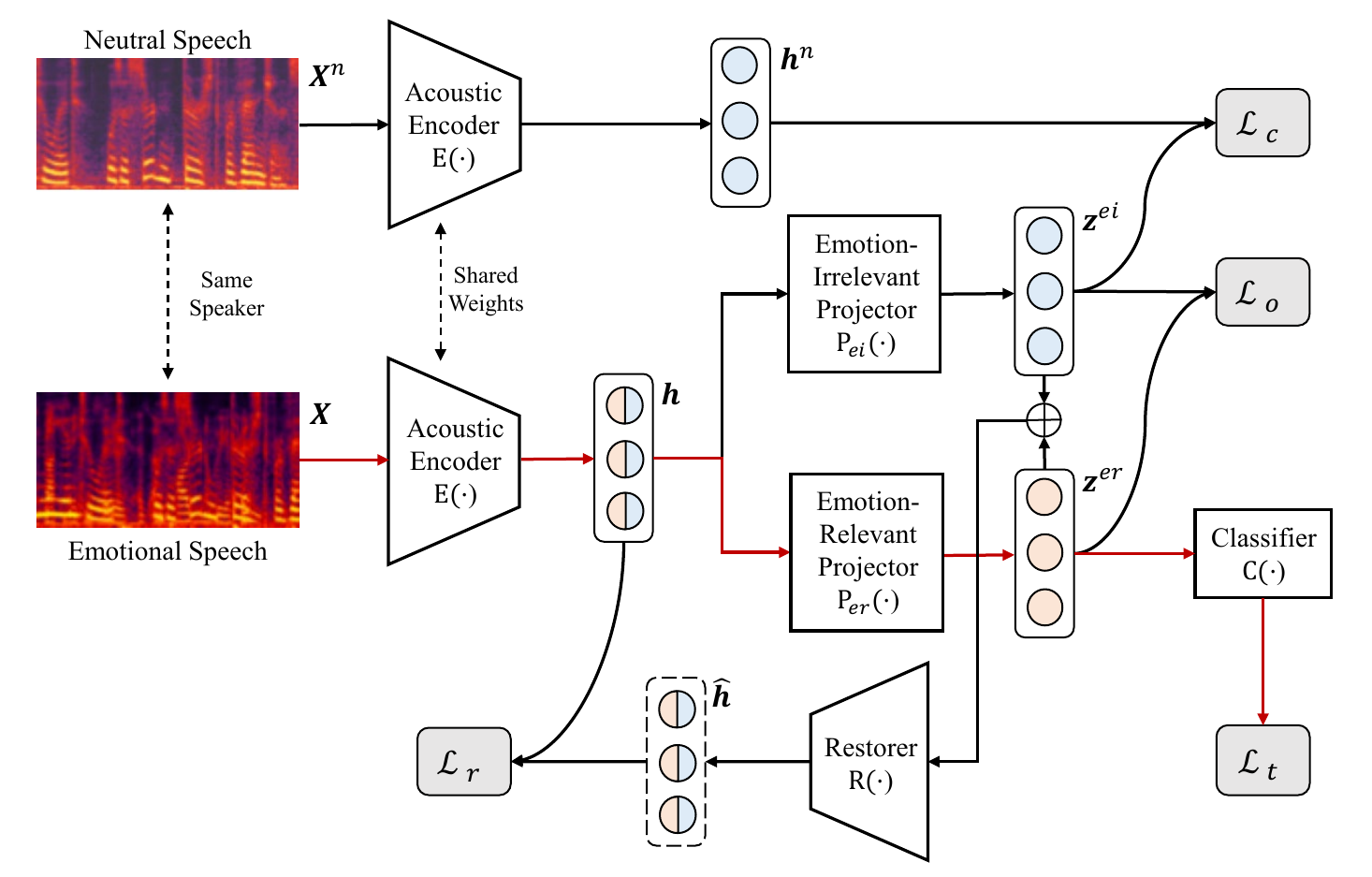}}
  \caption{The diagram of our proposed DSNet. The red arrows are activated throughout the training and inference phases, while the black arrows are activated only in the training phase.}
  \label{fig:overview}
\end{figure*}

As depicted in Fig.~\ref{fig:overview}, the overall workflow of our proposed DSNet consists of four major modules: an acoustic encoder $\mathrm{E}(\cdot)$ for feature abstraction, a pair of projectors $\mathrm{P}_{er}(\cdot)$ and $\mathrm{P}_{ei}(\cdot)$ for feature disentanglement, a restorer $\mathrm{R}(\cdot)$ for feature reconstruction, and a classifier $\mathrm{C}(\cdot)$ for emotion prediction.
In the following sub-sections, we present the details of each module.

\subsection{CNN-based Acoustic Encoder}
To start with, we denote the input speech spectrogram as $\boldsymbol{X} \in \mathbb{R}^{C \times T \times F}$, where $C$, $T$, and $F$ represent the dimensions of channel, time, and frequency, respectively.
The acoustic encoder is used to transform $\boldsymbol{X}$ into an utterance-level representation.
Within the encoder, multiple convolutional layers are first stacked with residual connection to extract the local characteristics of spectrogram.
Afterwards, average pooling is applied on the frequency axis, while attention pooling~\cite{attpool} is applied on the time axis for global information aggregation.
Formally, we have $\boldsymbol{h} = \mathrm{E}(\boldsymbol{X})$, where $\boldsymbol{h} \in \mathbb{R}^{D}$ denotes the deep feature vector with a fixed size of $D$.
Note that the encoder acts as the backbone network, which can be replaced by other architectures.

\subsection{Disentangled Siamese Network with Neutral Calibration}
Typically, the learned representation $\boldsymbol{h}$ is directly used as the input for an emotion classifier, and the ``black box'' is optimized in an end-to-end manner.
By contrast, our proposed DSNet aims to explicitly filter out the inadvertently encoded emotion-irrelevant information prior to the downstream emotion classification.

\subsubsection{Orthogonal Feature Disentanglement}
On the one hand, the orthogonal feature disentanglement module is designed to project $\boldsymbol{h}$ into independent feature subspaces.
Each projector is comprised of stacked attention blocks to focus on the specific elements of $\boldsymbol{h}$.
Concretely, the calculation within each attention block can be formulated as:
\begin{align}
\boldsymbol{a} &= \sigma \left( \operatorname{FC} \left(\delta \left( \operatorname{FC}\left(\boldsymbol{h}; \theta_1 \right) \right); \theta_2 \right) \right)\\
\boldsymbol{z} &= \delta \left( \operatorname{FC}\left(\boldsymbol{a} \circ \boldsymbol{h}+\boldsymbol{h}; \theta_3 \right)\right)
\end{align}
where $\theta_1, \theta_2, \theta_3$ are trainable parameters of the fully connected layers, and $\circ$ is the element-wise multiplication.
$\sigma$ and $\delta$ denote the sigmoid and ReLU activation functions, respectively.
We aim to disentangle $\boldsymbol{h}$ into the emotion-relevant and -irrelevant feature vectors via the following functions:
\begin{align}
\boldsymbol{z}^{er} &= \mathrm{P}_{er}(\boldsymbol{h}),\quad \boldsymbol{z}^{ei} = \mathrm{P}_{ei}(\boldsymbol{h})
\end{align}

Inspired by~\cite{dsn}, we impose a soft orthogonality constraint to encourage that $\boldsymbol{z}^{er}$ and $\boldsymbol{z}^{ei}$ encode different aspects of $\boldsymbol{h}$.
Let $\boldsymbol{Z}_{er} \in \mathbb{R}^{N \times D}$ and $\boldsymbol{Z}_{ei}\in \mathbb{R}^{N \times D}$ be the matrices whose rows are $\boldsymbol{z}^{er}$ and $\boldsymbol{z}^{ei}$, respectively, and $N$ denotes the batch size of the training set.
We first apply z-score normalization on the matrices, then the loss function can be calculated as:
\begin{align}
\mathcal{L}_{o} = \left \| \boldsymbol{Z}_{er}^\top \boldsymbol{Z}_{ei} \right \|^2_F
\end{align}
where $\left \| \cdot \right \|^2_F$ is the squared Frobenius Norm.

By enforcing the orthogonality constraint, the projectors have the risk of learning trivial solutions as they could produce orthogonal but uninformative vectors.
To this end, we impose an additional reconstruction constraint to ensure that $\boldsymbol{z}^{er}$ and $\boldsymbol{z}^{ei}$ retain the key information of $\boldsymbol{h}$.
We first obtain the reconstructed feature via an MLP-based restorer:
\begin{align}
\hat{\boldsymbol{h}} = \mathrm{R}(\operatorname{Concat} \left( \left[ \boldsymbol{z}^{er} ;  \boldsymbol{z}^{ei}\right] \right))
\end{align}
then the loss function can be calculated as:
\begin{align}
\mathcal{L}_{r} = \frac{1}{N} \sum_{i=1}^{N} \left \| \boldsymbol{h}_i - \hat{\boldsymbol{h}_i}\right \|^2_2
\end{align}
where $\left \| \cdot \right \|^2_2$ is the squared $L_2$ Norm.

\subsubsection{Siamese Neutral Calibration}
On the other hand, the Siamese neutral calibration module is designed to further direct the learning process for feature disentanglement.
Let $\boldsymbol{X}^n$ be a neutral spectrogram from the same speaker as the original input $\boldsymbol{X}$.
As shown in Fig.~\ref{fig:overview}, we also pass $\boldsymbol{X}^n$ through the encoder to get the deep feature vector $\boldsymbol{h}^n$.
We assume that $\boldsymbol{h}^n$ contains essential speech factors except for the emotional variation, then $\boldsymbol{h}^n$ can be utilized to guide $\boldsymbol{z}^{ei}$ in capturing emotion-irrelevant information.

Note that the Kullback Leibler (KL) divergence is a statistical metric to measure the discrepancy from one distribution to the other.
Mathematically, KL divergence is defined as:
\begin{align}
\mathrm{KL}\left( p,q \right) = \sum_{i=1}^{N}  p(x_i) \cdot  \log \frac {p(x_i)}{q(x_i)}
\end{align} 
where $p$ and $q$ are two probability distributions. 
Hence, the loss function of neutral calibration can be formulated as:
\begin{align}
\mathcal{L}_{c} = \frac{1}{2} \left( \mathrm{KL}\left( \boldsymbol{z}^{ei},\boldsymbol{h}^n \right) + \mathrm{KL}\left( \boldsymbol{h}^n,\boldsymbol{z}^{ei} \right)\right)
\end{align}

\subsection{Training Objectives}
The disentangled $\boldsymbol{z}^{er}$ is expected to concentrate on the emotion-relevant information of $\boldsymbol{h}$. 
Later, $\boldsymbol{z}^{er}$ is utilized as the input for an MLP-based emotion classifier, which can be formulated as $\hat{y}=\mathrm{C}(\boldsymbol{z}^{er})$.
The standard cross-entropy loss function is employed for the classification task:
\begin{align}
\mathcal{L}_{t} = - \frac{1}{N}\sum_{i=1}^{N} y_i\log{ \hat{y}_i}
\end{align}
where $y_i$ is the true label of the $i$-th sample in a training batch.
Finally, the overall loss function can be calculated as:
\begin{align}
\mathcal{L} = \mathcal{L}_t + \alpha \mathcal{L}_o + \beta \mathcal{L}_r + \gamma \mathcal{L}_c
\end{align}
where $\mathcal{L}_o$, $\mathcal{L}_r$, and $\mathcal{L}_c$ are loss functions of the orthogonality, reconstruction, and calibration constraints, respectively.
We empirically find that $\alpha=\beta=\gamma=1$ will suffice in our evaluation.
All the modules of DSNet are jointly optimized by minimizing $\mathcal{L}$.

\section{Experimental Datasets and Setup} \label{Experimental Datasets and Setup}
\subsection{Dataset Description}
To evaluate the performance of DSNet, we conduct experiments on two popular English datasets in SER, known as IEMOCAP~\cite{iemocap} and MSP-IMPROV~\cite{mspimprov}.

\textbf{IEMOCAP} contains five sessions of dyadic interactions between pairs of male-female actors in both scripted and improvised scenarios.
In order to maintain consistency with earlier studies, we merge the \emph{excited} category into the \emph{happy} category. As a result, the distribution of utterances used in the experiments is \{\emph{angry}: 1103, \emph{happy}: 1636, \emph{neutral}: 1708, \emph{sad}: 1084\}.

\textbf{MSP-IMPROV} contains six conversational sessions, where two actors interacted with each other in improvised scenarios, and one actor spoke the target sentences to make the emotion more natural. 
Four scenarios were developed for each target sentence to elicit different emotional responses. 
Finally, the distribution of utterances used in the experiments is \{\emph{angry}: 792, \emph{happy}: 2644, \emph{neutral}: 3477, \emph{sad}: 885\}.

\subsection{Implementation Details}
We first resampled the speech signal at 16 \emph{kHz}, then extracted the 80-dimensional Log Mel-scale Filter Bank (LMFB) with 25 \emph{ms} frame size and 10 \emph{ms} frame shift.
Following prior works, we used LMFB with deltas and delta-deltas as input acoustic features.
For parallel computing, the acoustic features of the same mini-batch were truncated or cyclically padded to a maximum length of 600 frames, and the batch size was set to 32.
We applied an Adam optimizer with a learning rate of $10^{-3}$ to optimize the model parameters.
The models were trained for a maximum of 100 epochs, and the learning rate was halved if the validation loss did not decrease for a consecutive 20 epochs.
We present the detailed configurations of our proposed DSNet in Table~\ref{tab:model_arch}.
The models were implemented using the PyTorch framework with NVIDIA Tesla P100 GPUs.
It should be noted that for each speaker, we chose one neutral reference $\boldsymbol{X}^n$ at random and kept it unchanged throughout model training.
Also, there was no overlap between $\boldsymbol{X}^n$ and $\boldsymbol{X}$. 

\begin{table}[htbp]
\centering
\caption{Detailed configurations of our proposed model. The size of input acoustic feature is $32 \times 3 \times 600 \times 80$.}
\label{tab:model_arch}
\begin{tabular}{l|l|l}
 \hline
 \textbf{Module} & \textbf{Settings} & \textbf{Output} \\
 \hline
 \multirow{8}{*}{Encoder}  & Conv2D(in=3,out=32,kernel=(5,5))+BN+ReLU &  $32 \times 32\times 600 \times 80$ \\
                          \cline{2-3} 
                          & MaxPool2D(kernel=(2,2)) &  $32 \times 32 \times 300 \times 40$               \\
                          \cline{2-3}
                          & Conv2D(in=32,out=64,kernel=(5,5))+BN+ReLU &  $32 \times 64\times 300 \times 40$ \\
                          \cline{2-3} 
                          & MaxPool2D(kernel=(2,2)) &  $32 \times 64\times 150 \times 20$               \\
                          \cline{2-3}
                          & Conv2D(in=64,out=128,kernel=(5,5))+BN+ReLU &  $32 \times 128 \times 150 \times 20$ \\
                          \cline{2-3}
                          & MaxPool2D(kernel=(2,2))   &  $32 \times 128\times 75 \times 10$             \\
                          \cline{2-3}                      
                          & Conv2D(in=128,out=256,kernel=(5,5))+BN+ReLU &  $32 \times 256 \times 150 \times 20$ \\
                          \cline{2-3}
                          & MaxPool2D(kernel=(2,2))   &  $32 \times 256 \times 37 \times 5$             \\
                          \cline{2-3}                            
                          & Frequency-wise Average Pool  &  $32 \times 256 \times 37$  \\
                          \cline{2-3}                       
                          & Temporal-wise Attention Pool  &  $32 \times 256$  \\                          
 \hline
  \multirow{4}{*}{Projector} & Linear(256,64)+ReLU+Linear(64,256)+Sigmoid &  \multirow{2}{*}{$32 \times 256$}  \\
                            & Linear(256,256)+ReLU \\                               
                            \cline{2-3}  
                            & Linear(256,64)+ReLU+Linear(64,256)+Sigmoid &  \multirow{2}{*}{$32 \times 256$}  \\
                            & Linear(256,256)+ReLU \\                          
 \hline
 Restorer                   & Linear(512,256)+ReLU+Linear(256,256)  &  $32 \times 256$  \\
 \hline
 \multirow{2}{*}{Classifier} & Linear(256,64)+BN+ReLU+Dropout &  $32 \times 64$  \\
                             \cline{2-3}
                             & Linear(64,4)+Softmax &  $32 \times 4$ \\  
 \hline
\end{tabular}
\end{table}

\subsection{Evaluation Settings}
We conducted both within-corpus (denoted as IEM and MSP) and cross-corpus (denoted as IEM2MSP and MSP2IEM) experiments to comprehensively evaluate the models in the speaker-independent scenarios. 
For within-corpus SER, leave-one-speaker-out cross-validation (LOSO CV) was employed to take full use of the limited data. 
Consider IEMOCAP as an example, a 10-fold LOSO CV was conducted, where 4 sessions were used for training, while utterances from the remaining two speakers were used for validating and testing, respectively.
The predictions and labels from each fold were concatenated to determine the final evaluation results.
Similar to this, a 12-fold LOSO CV was performed on MSP-IMPROV. 
While for cross-corpus SER, we directly evaluated the top models developed from the within-corpus experiments on the unseen dataset.
For instance, we evaluated the best 10 models trained on IEMOCAP using all the data from MSP-IMRPOV and averaged the results, and vice versa.
As for performance evaluation, we used the officially recommended metric, named unweighted average recall (UAR). UAR is defined as the arithmetic mean recall of all categories, which is insensitive to the impact of class imbalance.

\subsection{Comparative Methods}
To compare with our proposed method, we reproduced various state-of-the-art models that focused on the problem of cross-speaker generalization in SER. 
For a fair comparison, the input acoustic features and backbone networks were kept in line with our proposed DSNet. We briefly introduce these methods as below.

\textbf{MEnAN:} 
The Max-Entropy Adversarial Network (MEnAN)~\cite{debias-speaker} employs an iterative adversarial training strategy to optimize a multi-task framework.
While emotion-related information is maximized, speaker-related information is diminished by maximizing the entropy of the domain classifier output.

\textbf{DSC:} 
The Deep Speaker Conditioning (DSC)~\cite{dsc} introduces an auxiliary sub-network that conditions the primary classification network on a single neutral speech sample of the target speaker. To realize speaker adaptation, the reference samples are required in both the training and inference phases.

\textbf{ISNet:} 
The Individual Standardization Network (ISNet)~\cite{isnet} designs an independent encoder to generate individual benchmarks, which are used to standardize the emotional representations via the subtraction operation. Several neutral speech samples from the same individuals are required during the multistage training pipeline.

\textbf{DIFL:}
The Domain Invariant Feature Learning (DIFL)~\cite{difl} considers the speaker-independent SER problem as multi-source unsupervised domain adaptation.
The hierarchical alignment layer and multiple domain discriminators are proposed to jointly 
minimize differences across domains and confuse domain information for a domain-invariant emotional speech representation.

\section{Experimental Results and Analysis} \label{Experimental Results and Analysis}
\subsection{Comparison With Existing Works}
\subsubsection{Classification Performance Analysis} 
To start with, we perform the quantitative experiments and evaluate different approaches on both the IEMOCAP and MSP-IMPROV datasets.
According to the results presented in Table~\ref{tab:compare_table}, we conclude the following observations:

\begin{table}[htbp]
\centering
\caption{Performance comparison of different methods. Five random seeds were used to train each model, and the averaged results are reported. Speaker adaptation (SA) means entire or partial target utterances with speaker labels are incorporated in the training or inference phases. The best results are highlighted in bold.}
\label{tab:compare_table}
\begin{tabular}{l p{1.1cm}<{\centering} p{1.0cm}<{\centering} p{2.0cm}<{\centering} p{2.0cm}<{\centering} p{2.0cm}<{\centering} p{2.0cm}<{\centering}}
\toprule
 \multirow{2}{*}{Methods}  & \multirow{2}{*}{Year}  & \multirow{2}{*}{SA}  & \multicolumn{2}{c}{Within-corpus UAR(\%)} & \multicolumn{2}{c}{Cross-corpus UAR(\%)} \\
\cmidrule(lr){4-5}\cmidrule(lr){6-7}
           &     &       &   IEM             & MSP             & MSP2IEM             & IEM2MSP  \\
\midrule 
Baseline* & - &  \XSolidBrush &  56.13 $\pm$ 1.32  &  45.96 $\pm$ 2.38   &  38.24 $\pm$ 0.97   &  38.27 $\pm$ 0.88 \\
\midrule 
MEnAN & 2020 & \XSolidBrush   &  57.73 $\pm$ 0.98  &  47.70 $\pm$ 1.41   &  39.02 $\pm$ 1.34   &  38.15 $\pm$ 0.83 \\
DSC & 2021 & \Checkmark       &  57.29 $\pm$ 0.90  &  46.90 $\pm$ 0.59   &  38.74 $\pm$ 1.24   &  37.46 $\pm$ 0.26 \\  
ISNet & 2022 & \XSolidBrush   &  58.34 $\pm$ 1.31  &  48.50 $\pm$ 1.28   &  39.44 $\pm$ 1.23   &  37.25 $\pm$ 1.23 \\  
DIFL & 2022 & \Checkmark      &  58.07 $\pm$ 0.58  &  49.58 $\pm$ 0.62   &  39.09 $\pm$ 0.79   &  38.66 $\pm$ 0.41 \\  
\midrule 
DSNet  & 2023 & \XSolidBrush  &  \textbf{58.82} $\pm$ 0.70  &  \textbf{51.48} $\pm$ 0.87  &  \textbf{41.78} $\pm$ 0.51   &  \textbf{39.73} $\pm$ 0.99 \\   
\bottomrule 
\end{tabular}
\begin{tablenotes}
\scriptsize
\item[] *Baseline refers to the basic model with only acoustic encoder and classifier.
\end{tablenotes}
\end{table}

1) In the within-corpus settings, all the comparative methods remarkably outperform the baseline model on the two datasets, highlighting the efficacy of cross-speaker generalization for speaker-independent SER.
Moreover, our proposed DSNet exceeds the best performing comparative methods by 0.48\% and 1.90\% UAR on the IEMOCAP and MSP-IMPROV datasets, respectively.
We attribute these encouraging results to the explicit feature disentanglement in DSNet, which can better isolate and emphasize the emotion-relevant information within speech signals.

2) In the cross-corpus settings, however, most comparative methods can only achieve a slight performance gain over the baseline model.
In some cases, they can even perform worse than the baseline model.
While these techniques may be able to mitigate individual biases to some degree, we conjecture that they will inevitably learn the corpus-related biases.
By contrast, our proposed DSNet can surpass the baseline model by a significant margin of 3.54\% and 1.46\% UAR in the two cross-corpus scenarios, respectively.
We believe that the Siamese neutral calibration in DSNet can automatically discover the latent emotion-irrelevant factors including the corpus-related biases.

\subsubsection{Computational Complexity Analysis}
Subsequently, we compare the computational complexity of our proposed DSNet along with two other models that incorporate neutral speech samples in the training phase.
Specifically, we employ the metrics of model parameters and multiply-accumulate operations (MACs)~\footnote{\url{https://pypi.org/project/thop/}}.
Note that only partial modules of ISNet and DSNet are activated in the inference phase.
We also provide the computational complexity of the baseline model for reference purposes.

As shown in Table~\ref{tab:complexity}, we can observe that 
ISNet confronts notably higher MACs in the training phase due to the complex multi-stage model optimization, while DSC needs to process a pair of emotional and neutral samples in the inference phase, hence the parameters and MACs are doubled.
By contrast, the backbone network of DSNet is shared across the emotional and neutral samples in the training phase, and all the modules of DSNet are optimized jointly.
As a result, DSNet has the least model parameters and MACs in both the training and inference phases.

\begin{table}[htbp]
\centering
\caption{Computational complexity of different models.
The model parameters and MACs in the training and inference phases are reported, respectively.}
\label{tab:complexity}
\begin{tabular}{p{2.4cm} p{2.0cm}<{\centering} p{2.0cm}<{\centering} p{2.0cm}<{\centering} p{2.0cm}<{\centering}}
\toprule
{\multirow{2}{*}{Methods}} & \multicolumn{2}{c}{Model for Training}& \multicolumn{2}{c}{Model for Inference}\\
\cmidrule(lr){2-3} \cmidrule(lr){4-5} 
                 & Params (M)  & MACs (G) & Params (M)  & MACs (G)    \\
\midrule                        
Baseline            & 1.23  & 1.98 & 1.23  & 1.98  \\
\midrule 
DSC                 & 2.44 & 3.95 & 2.44 & 3.95  \\
ISNet               & 2.56 & 17.78 & 1.33 & 1.98  \\
DSNet (ours)        & 1.63 & 3.95 & 1.33 & 1.98  \\                
\bottomrule
\end{tabular}
\end{table}

\begin{figure*}[htb]
  \centering
  \scalebox{0.8}
  {\includegraphics[width=\linewidth]{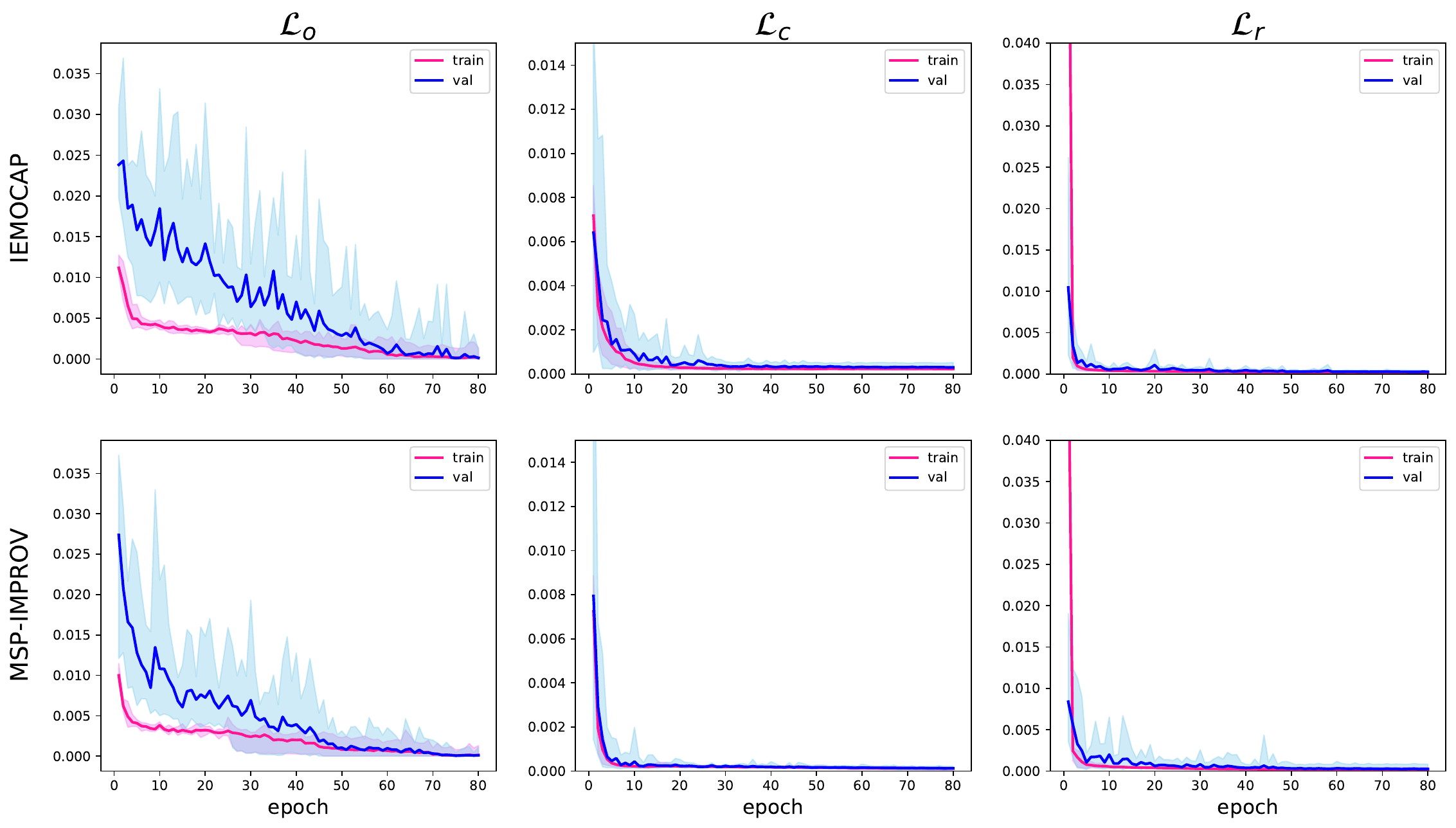}}
  \caption{Convergence curves of different regularization loss functions as the training progresses. The averaged values from LOSO CV on the IEMOCAP and MSP-IMPROV datasets are shown, respectvely. Additionally, the regions around each curve display the range of loss values across different experimental folds.}
  \label{fig:loss}
\end{figure*}

\subsection{Ablation Study}
In this subsection, we investigate the roles of different loss functions (\emph{i.e.}, $\mathcal{L}_o$, $\mathcal{L}_c$, and $\mathcal{L}_r$) for feature disentanglement.
Firstly, we plot the curves of loss variations with respect to the training epochs in Fig.~\ref{fig:loss}.
We can observe that both the training and validation loss values of $\mathcal{L}_c$ and $\mathcal{L}_r$ rapidly decrease in the first few epochs, while $\mathcal{L}_o$ is on a comparatively mild decline trend.
On both training and validation data, all of the loss functions eventually converge to an incredibly small value, indicating that the model is optimized in accordance with our expectations.

\begin{table}[htbp]
\centering
\caption{Ablation results of the constraints imposed on the feature disentanglement.}
\label{tab:ablate_table}
\begin{tabular}{p{2.4cm} p{1.8cm}<{\centering} p{1.8cm}<{\centering} p{1.8cm}<{\centering} p{1.8cm}<{\centering}}
\toprule
 \multirow{2}{*}{Model}   & \multicolumn{2}{c}{Within-corpus UAR(\%)} & \multicolumn{2}{c}{Cross-corpus UAR(\%)} \\
\cmidrule(lr){2-3}\cmidrule(lr){4-5}
               &   IEM             & MSP             & MSP2IEM             & IEM2MSP  \\
\midrule 
DSNet &  58.82  &  51.48   &  41.78   & 39.73  \\
\midrule 
DSNet w/o $\mathcal{L}_c$  &  57.60  &  49.73   &  40.53   &  36.08 \\
DSNet w/o $\mathcal{L}_o$  &  56.85  &  47.98   &  41.19   &  38.92 \\  
DSNet w/o $\mathcal{L}_r$  &  58.54  &  50.17   &  41.51   &  38.87 \\   
\bottomrule 
\end{tabular}
\end{table}

To verify the contributions of each loss function to the model robustness, we implement the following three variants of DSNet for comparison:
\begin{itemize}
\item DSNet w/o $\mathcal{L}_c$: It comes from DSNet but omits the calibration constraint.
\item DSNet w/o $\mathcal{L}_o$: It comes from DSNet but omits the orthogonality constraint.
\item DSNet w/o $\mathcal{L}_r$: It comes from DSNet but omits the reconstruction constraint.
\end{itemize}
The experimental results of different models are shown in Table~\ref{tab:ablate_table}.
We observe that all the three variants demonstrate inferior performance to the original DSNet, proving that each loss has positive impact on the model robustness.
More specifically, the removal of $\mathcal{L}_o$ causes the biggest performance degradation in the within-corpus settings, while the removal of $\mathcal{L}_c$ results in the greatest performance decline in the cross-corpus settings.
We argue that $\mathcal{L}_o$ is an important constraint to ensure the complementarity of the disentagled subspaces, while $\mathcal{L}_c$ plays a critical role in calibrating the emotion-irrelevant subspace especially in the cross-corpus settings.

\begin{figure*}[htb]
  \centering
  \scalebox{0.9}
  {
  \includegraphics[width=\linewidth]{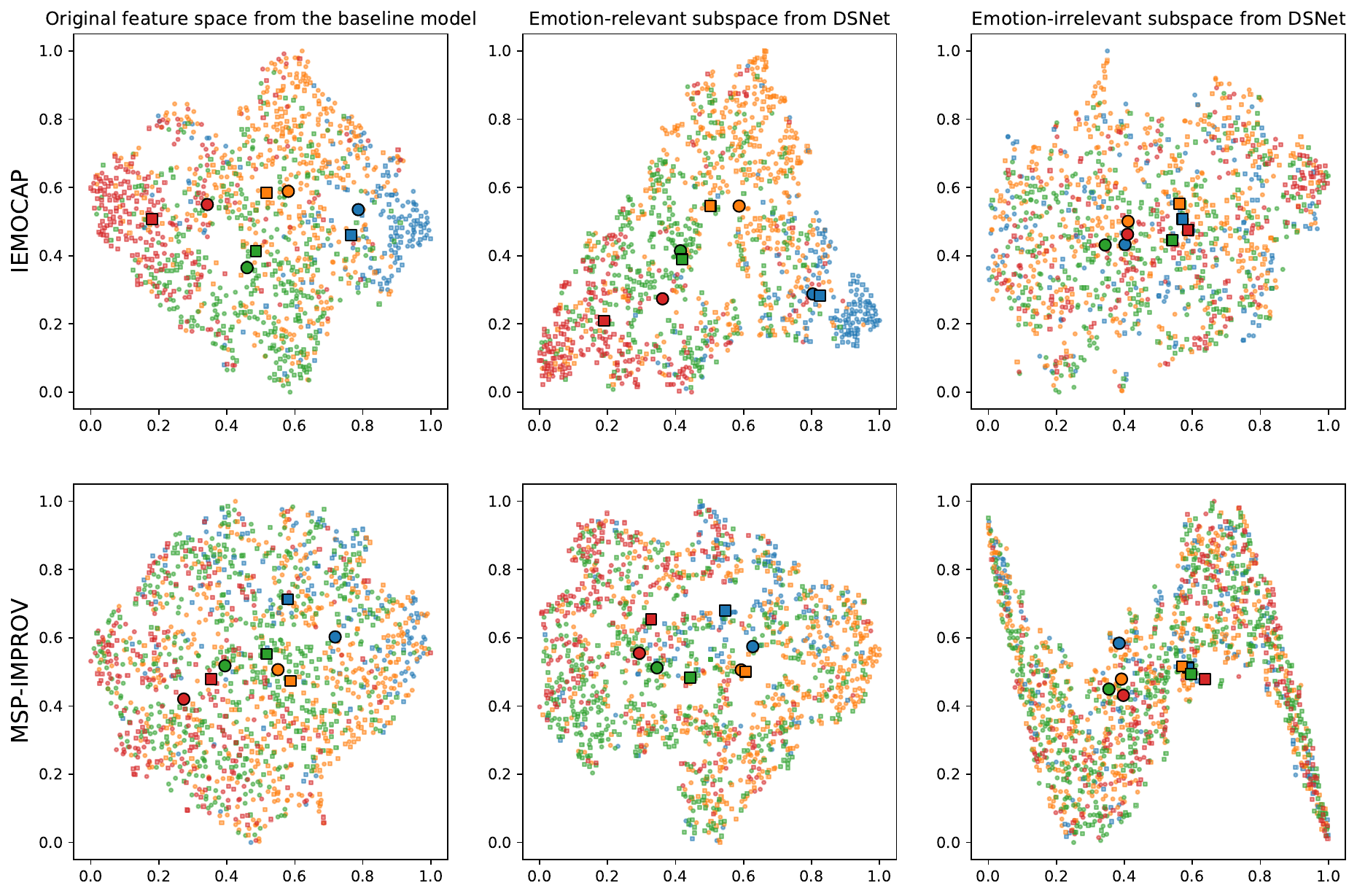}}
  \caption{T-SNE visualization for different feature spaces on the IEMOCAP and MSP-IMPROV datasets, respectively. In each subplot, the colors of blue, orange, green, and red denote the categories of \emph{angry}, \emph{happy}, \emph{neutral} and \emph{sad}, while the shapes of circle and square represent the validation and testing sets, respectively. The smaller semi-transparent points correspond to certain instances, while the larger opaque ones are the centers of each distribution.
  }
  \label{fig:visualize_ebd}
\end{figure*}

\subsection{Visualization Analysis}
In this subsection, we conduct the following visualization experiments to qualitatively analyze the efficacy of our proposed method.

\subsubsection{Visualization of Feature Spaces}
Fig.~\ref{fig:visualize_ebd} visualizes the distribution of original and disentangled feature spaces using t-SNE~\cite{tsne}.
Without loss of generality, we perform analysis on the latent representations of the validation and testing sets from the last fold.
According to the results displayed in Fig.~\ref{fig:visualize_ebd}, we conclude the following observations:

1) In contrast to the original feature space, the distance between the centers of the diverse emotional categories increases in the emotion-relevant subspace.
Meanwhile, the centers of different speakers belonging to the same emotional category are closer in the emotion-relevant subspace.
Consequently, the emotion-relevant subspace enjoys stronger emotional discrimination and cross-speaker robustness.

2) In the emotion-irrelevant subspace, however, the centers of various emotional categories from the same speaker tend to gather together.
It is conceivable that this subspace leaks little emotional information and can effectively discover speaker-related biases even on unseen data.

\begin{figure*}[htb]
  \centering
  \scalebox{0.78}
  {\includegraphics[width=\linewidth]{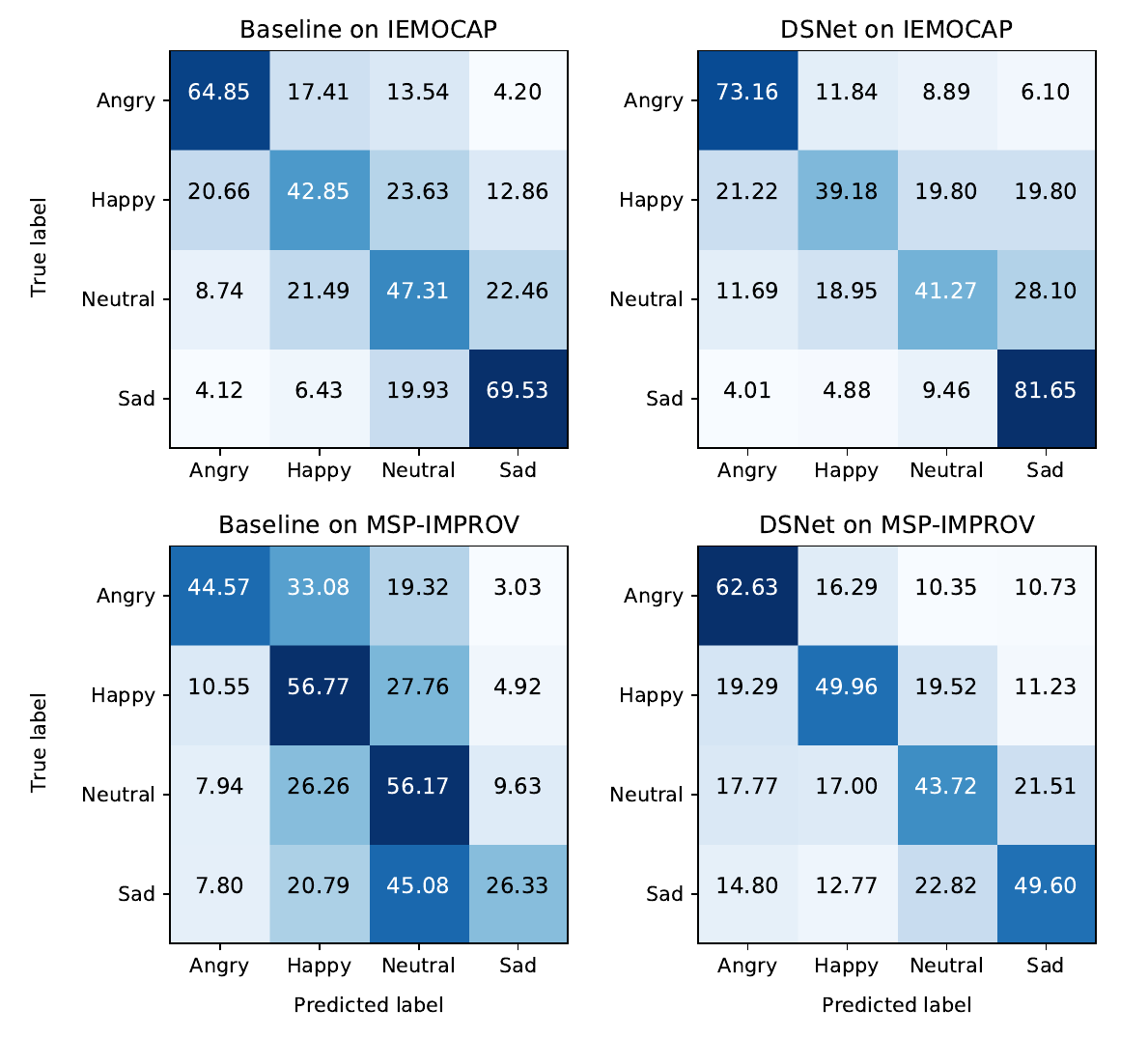}}
  \caption{Visualization for confusion matrices of the baseline model and our proposed DSNet on the IEMOCAP and MSP-IMPROV datasets, respectively.}
  \label{fig:visualize_cm}
\end{figure*}

\subsubsection{Visualization of Confusion Matrices}
Fig.~\ref{fig:visualize_cm} visualizes the confusion matrices of the baseline model and DSNet on the two datasets.
For the baseline model, we empirically find that \emph{happy} and \emph{neutral} are most likely to be confused with other categories, especially on the MSP-IMPROV dataset.
We speculate that the preponderance of data in these two categories may be the primary cause of this phenomenon.
In comparison, DSNet can better alleviate the phenomenon by filtering out latent emotion-irrelevant biases.
Moreover, DSNet is able to significantly improve recall of \emph{angry} and \emph{sad} by 8.31\%/18.06\% and 12.12\%/23.27\% on IEMOCAP/MSP-IMPROV, respectively.
As a result, DSNet consistently outperforms the baseline model in terms of UAR on both datasets.

\section{Conclusions} \label{Conclusions}
In this paper, we present DSNet, a simple yet effective framework that explicitly disentangles the high-level representation into emotion-relevant and -irrelevant components.
To guarantee the complementarity of disentangled feature subspaces, we incorporate a variety of loss functions as regularization, and jointly optimize all the modules of DSNet.
Quantitative experimental results demonstrate that DSNet is superior to currently advanced methods in terms of both classification performance and computational complexity.
We further verify the roles of different loss functions and the characteristics of disentangled subspaces via ablation study and visualization analysis.
Overall, DSNet can be regarded as a promising approach towards a more robust and explainable SER.
Future work will investigate the impact of neutral speech selection on the quality of disentangled subspaces.
Additionally, we intend to verify the effectiveness of our method on other backbone architectures.

\subsubsection{Acknowledgements} This research was partially supported by the National Key Research and Development Program of China (No. 2021YFC3320103).

%
%
%
%

\end{document}